\documentclass[aps,prl,reprint]{revtex4-2}
\usepackage{graphicx}
\usepackage{amsmath}
\usepackage{amssymb}
\usepackage{hyperref}
\begin{document}
	\title{Alterferroicity with the seesaw type magnetoelectricity}
	\author{Ziwen Wang}
	\author{Shuai Dong}
	\email{Corresponding author. Email: sdong@seu.edu.cn}
	\affiliation{School of Physics, Southeast University, Nanjing, China}
	\date{\today}
	\begin{abstract}
		Multiferroics provide a route to magnetoelectricity. But the general trade-off between magnetism and polarity remains inevitable. Here an alternative strategy is revealed, i.e., alterferroicity, which exhibits multiple but non-coexisting ferroic orders. Then the exclusion between magnetism and polarity becomes distinct superiority for strong magnetoelectricity. The design rules for alterferroics rely on the competition between the instabilities of phononic and electronic structures in covalent systems, like Ti-based trichalcogenides, which exhibit seesaw type magnetoelectricity and possible ferroic phase separation.
	\end{abstract}
	\maketitle

\begin{figure*}
\includegraphics[width=\textwidth]{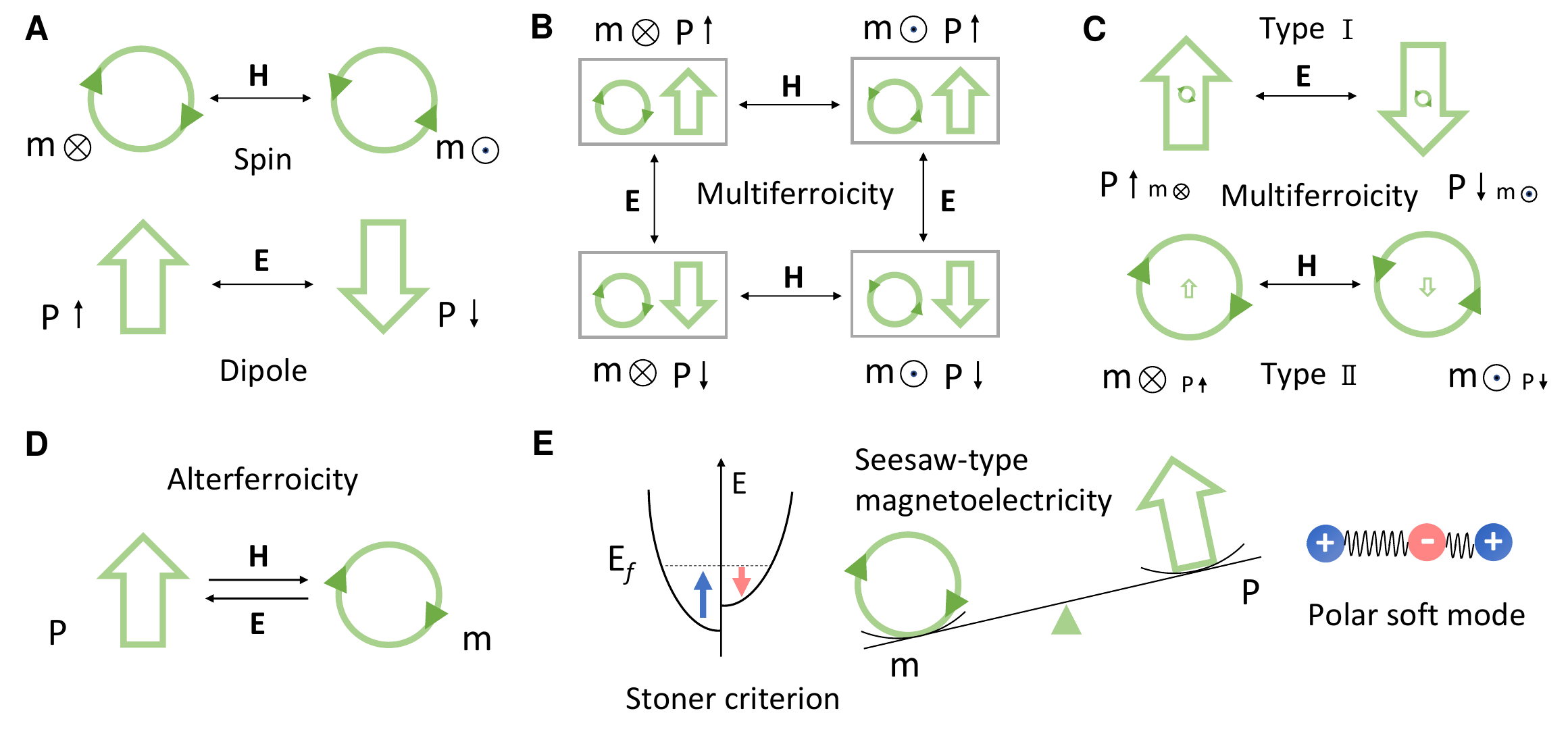}
\caption{Schematic of ferroicities: from primary ferroicities, to multiferroicity, then to alterferroicity. (\textit{A}) Two most popular primary ferroicities. Spin moments can be switched by magnetic fields. Dipole moments can be switched by electric fields. (\textit{B}) General scenario of multiferroicity with coexisting magnetism and polarity. The magnetic/electric fields can only directly switch the spins/dipoles correspondingly, while their cross controls remain tough. (\textit{C}) The subtle magnetoelectric cross controls in the type-I and type-II multiferroic systems. However, here the net magnetizations in the type-I case and dipoles in the type-II case are both small secondary quantities, instead of primary order parameters. (\textit{D}) The concept of alterferroicity, with multiple exclusive ferroic orders, which allow the direct switch between the primary magnetism and polarity. (\textit{E}) Origin of alterferroicity: competition between the electronic instability (left) and phononic instability (right). Middle: schematic of seesaw type magnetoelectricy: an electric (magnetic) field can tune the phase balance between magnetic and polar state, thus the magnetism (polarization) can be modulated nonlinearly.}
	\label{F1}
\end{figure*}

\textit{Introduction}. 
Ferroic quantities, e.g. magnetization or polarization, can be tuned by their conjugate fields (magnetic field or electric field) as shown in Fig.~\ref{F1}\textit {A}, which provide fundamental functions for many device applications. As a hybrid branch of ferroicities, multiferroicity is expected to host intrinsic magnetoelectricity, namely to control magnetism via electric field, or control polarization via magnetic field, which will be highly valuable for next-generation information devices \cite{Eerenstein2006,Cheong:Nm,Dong2015,Fiebig2016,Spaldin2019}. 

However, there is a fundamental paradox along this route, which prohibits the practical success of magnetoelectric cross switching. Generally, proper polarity and magnetism are naturally untwisted (or even exclusive) from the quantum electronic level. Even if one can find an ``ideal'' multiferroic system with both proper ferromagnetism and ferroelectricity, it would remain tough to obtain the cross-switching functions since these two ferroic orders obey different symmetries \cite{Dong2015}, as sketched in Fig.~\ref{F1}\textit{B}.  

The progress of multiferroicity in past decades has found some tricks to circumvent this paradox \cite{Dong2019}, as summarized in Fig.~\ref{F1}\textit{C}. For example, in some type-I multiferroics, the spin chiralities (and associated weak ferromagnetic moments) are triggered by the polarizations, via the spin canting of antiferromagnetic background or at domain walls \cite{Heron2014,Chen2021}. In contrast, in the so-called type-II multiferroics, the tiny ferroelectric dipoles can stem from special magnetic textures \cite{Kimura2003,Katsura:Prl,Zhang:Jacs}, not as the primary order parameter. Despite these achievements, limited by the fundamental principles, all these solutions are based on the secondary effects, which are naturally faint and inadequate to be implemented in practice.

\textit{Design rules of alterferroicity}.
Alterferroicity, defined as a switchable hybrid ferroicity between multiple primary ferroicities, is a sister branch of multiferroicity, which can tell a totally different story, as sketched in Fig.~\ref{F1}\textit{D}. On one hand, similar to multiferroicity, there are multiple (usually two) primary ferroic orders existing in an alterferroic material. On the other hand, these ferroic orders are not simultaneously coexisting in the same phase as in multiferroics. In other words, in alterferroics, we no longer seek the awkward cooperation of magnetism and polarity, but take full advantage of their natural exclusion. Then, by tuning the subtle balance between the mutually exclusive magnetic and polar phases, one can switch on/off magnetism (polarization) using electric (magnetic) fields, as sketched in Fig.~\ref{F1}\textit{D-E}. This intrinisc strong magnetoelectricity is conceptually different from conventional linear magnetoelectricity nor nonlinear magnetoelectricity in those type-II multiferroics, which is coined as the seesaw-type magnetoelectricity.

Just like the correspondences between physical conversations and their conjugate symmetries, in general, there are correspondences in solids beween primary ferroic orders and certain instabilities. Specifically, polarity in crystals originates from the dynamic instability of crsytalline structure, i.e., polar soft modes of phonons, while magnetism in solids comes from the instability of electronic structure, i.e. Stoner criterion, as shown in Fig.~\ref{F1}\textit{E}. In this sense, the first design rule of alterferroicity is to find candidates with dual instabilities, which should be comparable in intensity and mutually exclusive.

To realize alterferroicity, those systems near the boundary of polarity and magnetism are desired. Taking the mostly-studied transition metal compounds for example, generally, magnetic moments usually originate from unpaired $d$ electrons, while dipoles are generated by the coordinate bonding between empty $d$ orbitals and neighoring close-shell $p$ orbitals. These rules have been known for decades \cite{Hill2000}. Historically, this scenario was established mostly based on ionic crystals. In contrast, in covalent systems the chemical valences of ions are not so rigid but relatively ``soft'', which provides a tolerant environment to tune the subtle balance between polarity and magnetism. Thus, the second design rule for alterferroicity is to seek those candidates with (partial) covalent bondings between metals and anions. Thus, oxides and halides are excluded since the strong electronegativities of oxygen and halogens prefer the ionic bonding.

~\\
\textit{Alterferroic materials}.
Following these design rules, we implement the concept of alterferroicity in transition metal trichalcogenides (TMTC's) monolayer. With a general formula $ABC_3$ ($A$: metal; $B$: auxiliary ion usually pnicogen or group IV element; $C$: chalcogen except oxygen), TMTC's form a large family of van der Waals (vdW) layered crystals \cite{Ouvrard1988,Carteaux1995,Susner2017}. Both magnetic and polar TMTC's have been reported, such as ferroelectric CuInP$_2$S$_6$ \cite{Liu2016} and ferromagnetic CrGeTe$_3$ \cite{Gong2017}. Other antiferromagnets like $A$PS$_3$ ($A$=Fe, Mn, Ni) have also been well known \cite{Samal:Jmca,Lu:Jpcm}, and recently a few more ferroelectric candidates were predicted in some TMTC's monolayers by a high-throughput calculation \cite{Hao2021}. In addition, the weak electronegativities of chalcogens and auxiliary ions allow (partial) covalent bonds. The last but not the least, we choose Ti as the core ion $A$, which can be either dipole active or magnetic active, depending on its valence (Ti$^{4+}$ \textit{vs} Ti$^{3+}$).

\begin{table}
	\caption{Basic physical properties of the competing phases in TiGeTe$_3$ monolayer. All values are obtained in DFT calculations with default $U_{\rm eff}$(Ti)=$2.5$ eV. $a$ and $b$ are the lattice constants. The energies per f.u. are in relative to the parent phase.}
	\begin{tabular*}{0.48\textwidth}{@{\extracolsep{\fill}}lcccc}
		\hline \hline
		Phase & S.G. & $a$/$b$ (\AA) & Energy (meV) & Gap (eV) \\
		\hline
		Parent & $P\bar{3}1m$  & $6.952$/- & $0$ & $0$ \\
		Ferroelectric & $P31m$ & $7.071$/- & $-73$ & $0.22$ \\
		Z-AFM & $C2/m$   & $6.980$/$12.201$  & $-264$ & $0.46$ \\
		\hline \hline
	\end{tabular*}
	\label{Tab-1}
\end{table}

\begin{figure}
	\includegraphics[width=0.48\textwidth]{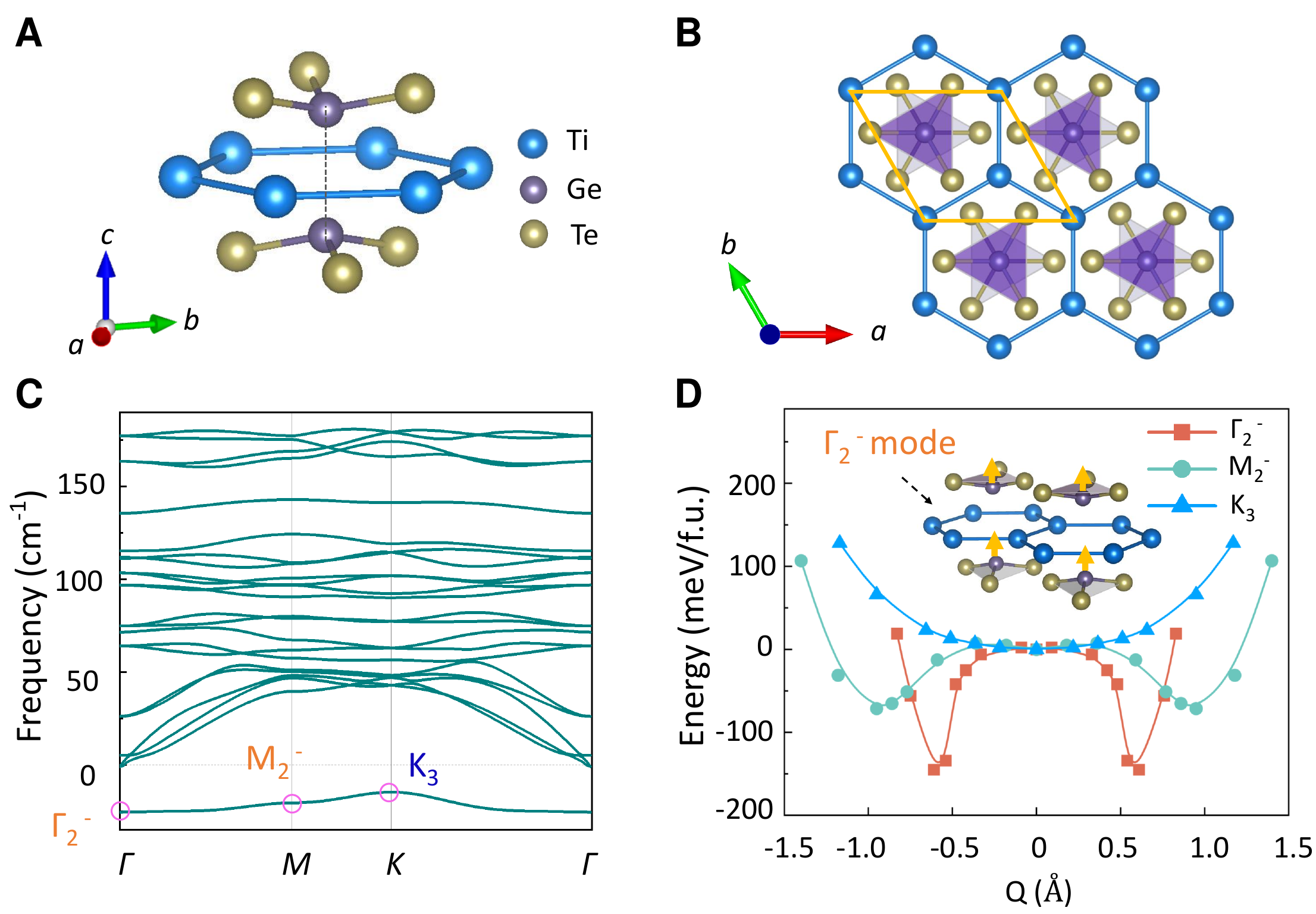}
	\caption{Structures and instability of crystalline of alterferroic TiGeTe$_3$ monolayer. (\textit{A}) Side view of a structural unit, consisting of sandwich layers. (\textit{B}) Top view. The primitive cell is denoted by the orange rhombus. (\textit{C}) The phonon spectrum of the high-symmetric ($P\bar{3}1m$) nonmagnetic phase, which contains a branch of significant imaginary frequency. Such an imaginary frequency will lead to spontaneous distortions. (\textit{D}) The energy profiles of corresponding distortions ($\Gamma_2^-$, $M_2^-$, and $K_3$) without magnetism. The $\Gamma_2^-$ mode is the most preferred one. Insert: schematic of atomic vibration of the $\Gamma_2^-$ mode, which is polar along the $c$-axis.}
	\label{F2}
\end{figure}

Our first candidate system is TiGeTe$_3$, which is isostructural with CrGeTe$_3$ monolayer \cite{Zhang2015,Gong2017}. As shown in Fig.~\ref{F2}\textit{A}, the middle layer is Ti-honeycomb, sandwiched by GeTe$_3$ layers. Its primitive cell is a rhombus, as depicted in Fig.~\ref{F2}\textit{B}, with in-plane triple rotational ($C_3$) symmetry.

To reveal the ferroicities of TiGeTe$_3$, primary first-principles density functional theory (DFT) calculations are performed. Its phonon spectrum of high-symmetric nonmagnetic phase (i.e., the parent phase) is shown in Fig.~\ref{F2}\textit{C}, which contains a branch of imaginary frequency over the whole Brillouin zone (BZ). The strongest imaginary frequency appears at the BZ center (the $\Gamma_2^-$ mode), while the $M_2^-$ and $K_3$ modes at the BZ boundary are slightly weaker. In particular, the $\Gamma_2^-$ mode leads to the ferroelectric distortion along the $c$-axis [inset of Fig.~\ref{F2}\textit{D}], while the $M_2^-$ and $K_3$ modes lead to antiferroelectric configurations as shown in SI Appendix, Fig.~S1.

Then the energy profiles of distortion modes are calculated, as shown in Fig.~\ref{F2}\textit{D}. It is clear that the $\Gamma_2^-$ mode dominates the polarity, with a much deeper energy well. In this sense, the polar ground state should be a ferroelectric one, instead of those antiferroelectric states. 

\begin{figure}
	\includegraphics[width=0.48\textwidth]{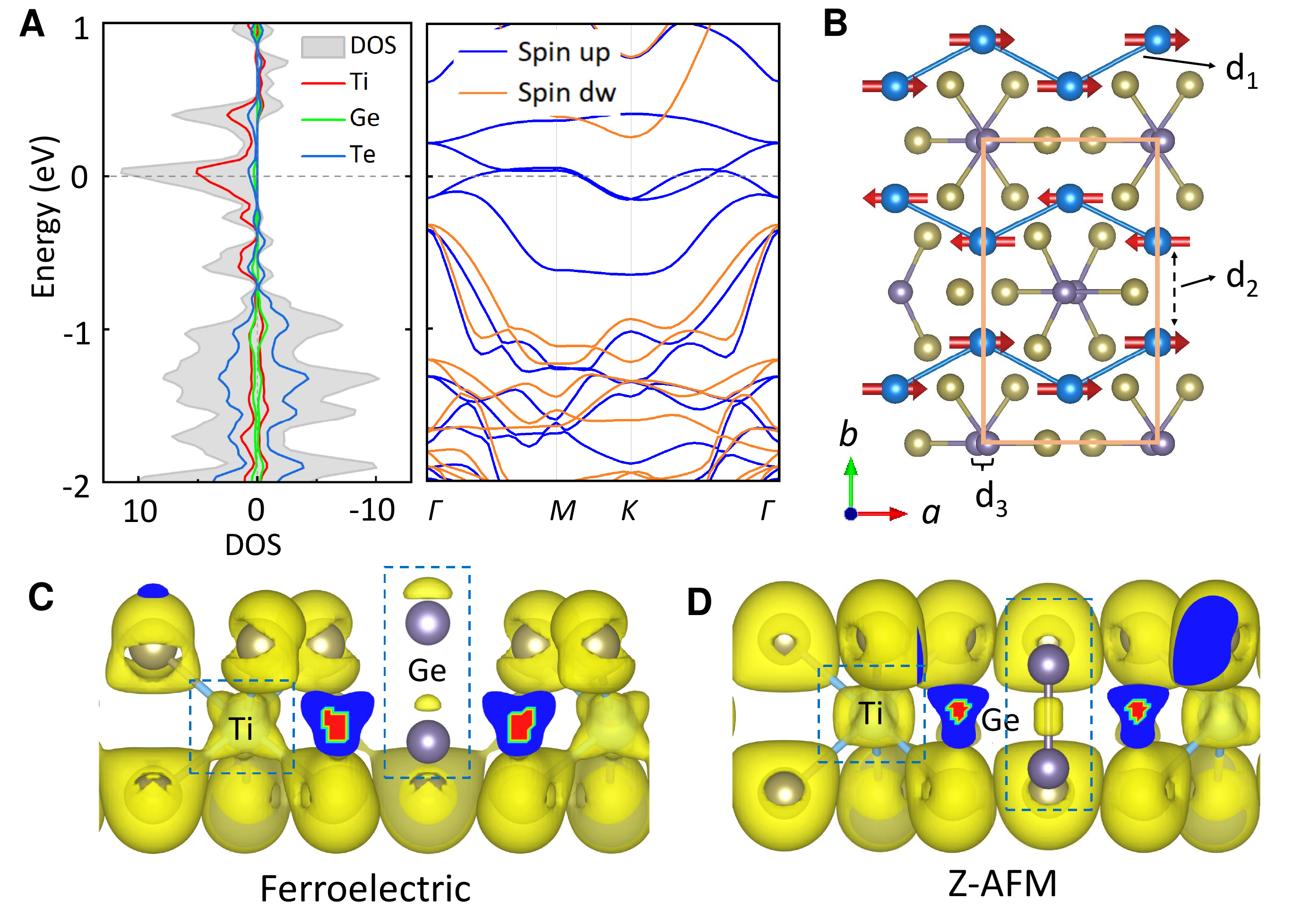}
	\caption{Instability of electronic structures of alterferroic TiGeTe$_3$ monolayer. (\textit{A}) Electronic structure of high-symmetric ferromagnetic phase. Left: total DOS and contributions from individual atomic species; Right: band structure. There is a peak of DOS from Ti at the Fermi level, implying the van Hove singularity. (\textit{B}) Structure of the orthorhombic Z-AFM phase. Arrows: spins of Ti; $d_1$ and $d_2$: the disproportions of Ti-Ti distances; $d_3$: the relative horizontal displacements between the upper/lower Ge ions; Orange rectangle: a primitive cell. Visualization of electronic clouds of top valence bands for comparison: (\textit{C}) The ferroelectric phase; (\textit{D}) The Z-AFM phase. The key differences are emphasized in boxes.}
	\label{F3}
\end{figure}

However, the ferroelectric state is not the whole story for TiGeTe$_3$. In analogy to Cr ion in CrGeTe$_3$, the nominal value of Ti ion in TiGeTe$_3$ is $+3$, which is magnetic active according to the emprical rule of oxides. Indeed, for the parent phase, its electronic density of states (DOS) shows a peak around the Fermi level (SI Appendix, Fig.~S2), which is a precondition of Stoner instability. Then the electronic structure of ferromagnetic state is also calculated, as shown in Fig.~\ref{F3}\textit{A}. A strong DOS peak remains very close to the Fermi level, implying a spin-polarized van Hove singularity, which can lead to the Slater-type metal-insulator transition to an antiferromagnetic state.

Here four most possible magnetic orders in the honeycomb lattice are considered, including ferromagnet (FM), N\'eel-type antiferromagnet (N-AFM), stripy-type antiferromagnet (S-AFM), and zigzag-type antiferromagnet (Z-AFM) (SI Appendix, Fig.~S3). Our calculation finds that the Z-AFM owns the lowest energy, as compared in SI Appendix, Tab.~S1. This Z-AFM breaks the $C_3$ symmetry of honeycomb lattice, i.e., changing the trigonal Bravais lattice to an orthorhombic one, as depicted in Fig.~\ref{F3}\textit{B}.

As summarized in Table~\ref{Tab-1}, the Z-AFM phase is centrosymmetric in crystal structure, with a magnetic moment $\sim0.88$ $\mu_{\rm B}$/Ti, as expected for Ti$^{3+}$. In contrast, the ferroelectric phase is nonmagnetic, with a net polarization $2.06$ pC/m along the $c$-axis. To check their mutual exclusion, hypothetic multiferroic phases, i.e., polar magnetic states, are also tested, but all magnetic moments quench to zero after the optimization.

It is necessary to check the stabilities of these competing phases. For both the optimized ferroelectric structure and orthorhombic Z-AFM structure, there is no obvious imaginary frequency anymore (SI Appendix, Fig.~S4). Therefore, both of them are dynamically stable. Their DOS's are also calculated (SI Appendix, Fig.~S5), both of which are insulating. Then the original electronic instabilities are also eliminated in both cases. In short, the parent phase of TiGeTe$_3$ monolayer is unstable, but its two derived ferroic phases are both stable.

To illustrate the electronic origin of alterferroicity, the electronic clouds of top valence electrons are compared in Figs.~\ref{F3}\textit{C-D}. The subtle but vital differences between the ferroelectric and Z-AFM states are twofold. First, the unilateral $d$-$p$ hybridizations of Ti-Te coordination bonds is evident in the ferroelectric state, but absent in the Z-AFM one. Second, in the ferroelectric phase, the $4s^2$ lone pairs of Ge ions can be seen as a unilateral ``hat'', while in the Z-AFM phase, it is covalent between the Ge-Ge pair. In this sense, here Ge ions also contribute significantly to the ferroelectricity. More comparisons of Ti-Te and Te-Ge bonds can be found in SI Appendix, Fig.~S6.

\begin{figure*}
	\includegraphics[width=\textwidth]{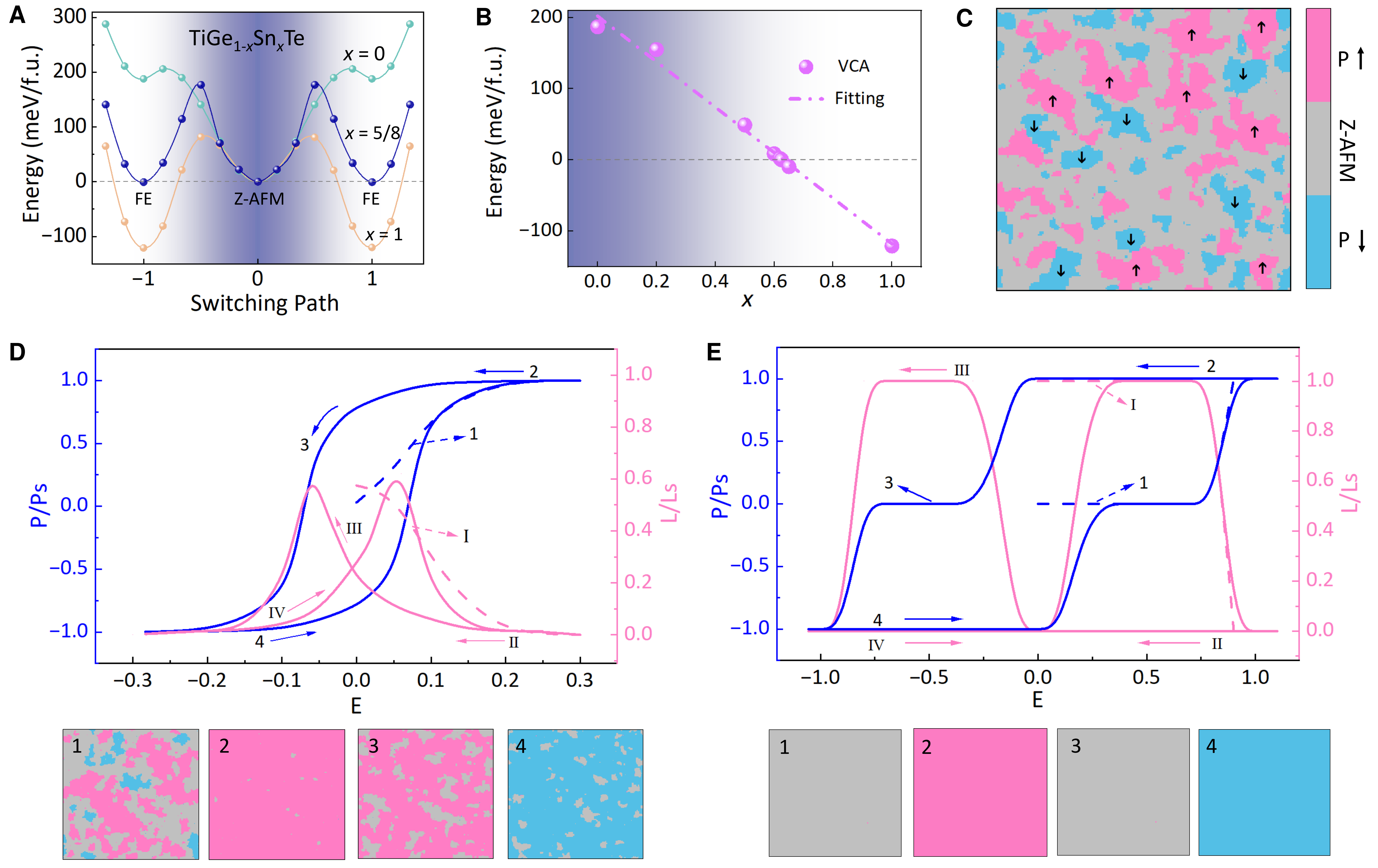}
	\centering
	\caption{Fine tuning of alterferroicity and the seesaw type magnetoelectricity. (\textit{A}) The DFT switching paths of alterferroic Ti(Ge$_{1-x}$Sn$_x$)Te$_3$ monolayers. (\textit{B}) The DFT energy difference between the ferroelectric and Z-AFM states as a function of Sn concentration $x$. With a proper $x=5/8$, the energies of ferroelectric and Z-AFM states can be exactly identical. (\textit{C}) A typical Monte Carlo snapshot of ferroic inhomogeneity based on the random-field pseudo-dipole model. (\textit{D-E}) Typical hysterisis loops of alterferroicity obtained by Monte Carlo simulation, as a function of electric field. Left axis: normalized net polarization; Right axis: volume weigth of Z-AFM component. The sequence 1-2-3-4 for the FE hysteresis loop (and the corresponding I-II-III-IV for the magnetic hysteresis loop) denotes the sweeping process of electric field, and the broken curves denote the initial process. Typical snapshots of corresponding segments are shown below the hysterisis loops. (\textit{D}) With the quenching disorder and equivalent energies of Z-AFM and FE phases, i.e., the $x=5/8$ case in (\textit{A}). (\textit{E}) Without the quenching disorder and unequivalent energies, i.e., the pure TiGeTe$_3$ case. In Monte Carlo snapshots, different phases are distinguished by colors, as depicted in the label of (\textit{C}). More details of model parameters can be found in SI Appendix.}
	\label{F4}
\end{figure*}

\textit{Seesaw type magnetoelectricity}.
Since both the ferroelectric and Z-AFM states are stable, in principle the ground state of TiGeTe$_3$ monolayer can be switched between the nonpolar-magnetic one and polar-nonmagnetic one using by external stimulus like electric field, realizing the goal of electric field switching on/off magnetism. This tuning process renders a seesaw like magnetoelectric function, which is a unique characteristic of alterferroicity, distinct from the established scenario of multiferroicity. The DFT energy profile of TiGeTe$_3$ for alterferroic switching is sketched in Fig.~\ref{F4}\textit{A}, as a function of normalized ferroelectric distortion. An energy barrier between Z-AFM and FE phase is in the order of $200$ meV/f.u.. For comparison, the most extensively studied multiferroic, BiFeO$_3$, exhibits a $0.43$ eV/f.u. theoretical energy barrier for the FE reversal~\cite{Ravindran2006}. Noting the switching paths in Fig.~\ref{F4}\textit{A} are obtained for a uniform transition, which can only be considered as a theoretical upper limit. In fact, other paths with lower barriers are also possible. For example, the switching can be done non-uniformly with the help of multiple domain structures, which can effectively reduce the electric field requirement (to be discussed later).

Although the concept of alterferroicity has been demonstrated, their roles remain not fully equivalent, since the Z-AFM is more energetically favourable ($191$ meV/f.u. lower than the competitor). Thus in the pristine TiGeTe$_3$ monolayer, the Z-AFM phase will be dominant while the ferroelectric state is hard to be obtained in practice. Then is it possible to realize an unbiased alterferroic system with energetic-equivalent ferroic states?

To fulfill this aim, other Ti-based TMTC's monolayers are studied by replacing Ge using other group IV elements. The calculated basic physical properties are shown in SI Appendix, Tab.~S2. First, for TiCTe$_3$ and TiSiTe$_3$, they do not exhibit the ferroelectricity, but only magnetism. Second, for TiSnTe$_3$ and TiPbTe$_3$, their ferroelectric states become energetically more favourable than magnetic states, in opposite to the Ge-case. This tendency is reasonable. First, larger ions dilate the structures, which is emprically advantageous for ferroelectric distortions. Second, the more spatially expanding $5s^2$ or $6s^2$ lone pairs of Sn/Pb can assist the ferroelectricity more.

Therefore, it must have a compensated point between TiGeTe$_3$ and TiSnTe$_3$, which owns the energetic-equivalent ferroic states. Thus we calculate the mixture cases of Ti(Ge$_{1-x}$Sn$_x$)Te$_3$. As shown in Figs.~\ref{F4}\textit{B}, the energy difference between the ferroelectric and Z-AFM states can be gradually tuned by the substitution $x$, which reaches zero at $x=5/8$, and the alterferroic switching profile at this compensated point (blue dotted solid line in Fig.~\ref{F4}\textit{A}) indeed shows wells with equivalent depth.

Above DFT calculations on Ti(Ge$_{1-x}$Sn$_x$)Te$_3$ are obtained in a mean-field level. To further demononstrate the seesaw type magnetoelectricity of alterferroicity with realistic ionic disorder, we construct a phenomenological random-field pseudo-dipole model with minimum components \cite{Mayr:Prl,Zhu:Nc}. The quench disorder introduced by Ge/Sn ionic mixture is mimicked by a random field, which can induce local fluctuations regarding the competitive ferroic phases. With such local fluctuations, the alterferroic system can exhibit ferroic inhomogeneity in the Monte Carlo simulation, as shown in Fig.~\ref{F4}\textit{C}. Phenomenally, this ferroic inhomogeneity is similar to the magnetoelectric composites \cite{Nan:Jap}. However, the underlying physics is essentially different from those composites, but more close to the phase separation in correlated electronic systems \cite{Dagotto:Sci}. The typical scale of alterferroic inhomogeneities is roughly estimated to be in the nanometer, as shown in SI Appendix, Fig.~S7.

By applying an external field (e.g. electric field), the ferroic phase separation pattern can be reshaped. Consequently, the macroscopic (average) ferroic quantities are naturally modulated, as shown in Fig.~\ref{F4}\textit{D}. When applying a positive electric field, the FE(P up) phase will be preferred and these clusters will grow larger and larger, while the FE(P down) clusters and Z-AFM clusters will shrink accordingly. Such reshaping process of inhomogeneity leads to the increasing of net polarization, i.e., the segment 1 of hysteresis loop in Fig.~\ref{F4}\textit{D}. In the high electric field limit, the system is (almost) fully ferroelectric, i.e., a single FE(P up) domain. Then with decreasing field to zero, i.e., the segment 2 of hysteresis loop, the single domain remains stable but a few small clusters of Z-AFM and FE(P down) will emerge due to thermal fluctuation. Furthermore, with increasing negative electric field (segment 3), the FE(P up) clusters will rapidly shrink. More and larger Z-AFM clusters will appear rapidly and reaches a peak near the coercive field. After that, the FE(P down) clusters will also increase and finally take over the whole lattice.

It should be noted that even though the energies of the two competing phases are not idential (e.g. for the pure TiGeTe$_3$ or small doping case), the transition energy barrier could also stabilize these phases in practice \cite{Ding2021}. For the pure TiGeTe$_3$ without quenching disorder, the energy of Z-AFM state is lower than the FE state, and the simulated hysterisis loops are shown in Fig.~\ref{F4}\textit{E}. There are two stable shoulders in the FE loop, corresponding to the pure Z-AFM phase in the low electric field region. Lacking the obvious phase separation, the larger coercive field is required during the ferroic switching.

Such seesaw type magnetoelectricity is a strong effect, with canonical switches between two primary ferroic orders. In principle, the magnetic switching should also be expectable, since the response to magnetic field (i.e., the susceptibility) could be different between the antiferromagnetic phase and nonmagnetic polar phase, although the antiferromagnetic phase is not as ideal as a ferromagnetic one.

Besides the electric field, strains can also be used as driving force to tune the alterferroicity. As shown in Fig.~\ref{F3}\textit{B} and Table~\ref{Tab-1}, the Z-AFM state has an orthorhombic type distortion. The lattice mismatch between competing ferroelectric and Z-AFM phases leaves the space to tune the alterferroic phase equilibrium and phase separation (SI Appendix, Fig.~S8). In fact, similar to the two polar value ($\pm P$) of ferroelectric phase, there are $\mathbb{Z}_3\times\mathbb{Z}_2$ antiferromagnetic values ($\mathbb{Z}_3$: three zig-zag orientations derived from the $C_3$ symmetry of honeycomb lattice; $\mathbb{Z}_2$: antifermagnetic order $\pm L$). Consequently, a large piezoelectric and piezomagnetic effects can be expected, and a rough estimation of the piezoelectric stress coefficient e$_{31}$ is about 400 pC/m. For a hypothetic alterferroic with the ferromagnetism, $\sim1$ $\mu_{\rm B}$/f.u. magnetization can be changed by $\sim2\%$ strain. These multiple degrees of freedom deserve future studies. 

Last but not the least, the concept of alterferroicity does not exclude the possibilities of other magnetic orders. For example, in a sister member TiGeSe$_3$, the competing magnetic phase becomes the S-AFM one (SI Appendix, Tab.~S3 and Fig.~S9). It will be more interesting to involve the ferromagnetism in alterferroicity for broader potential applications. For these alterferroic materials with ferromagnetic states, a semiquantitative estimation of the seesaw type magnetoelectricity is: an electric field of typical ferroelectric coercivity (e.g. $\sim10^4$-$10^5$ V/cm) can switch on/off $\sim1$ $\mu_{\rm B}$/f.u. magnetization.

\textit{Conclusions}.
Our study has proposed an emerging concept alterferroicity as the sister of multiferroicity. This previously missed ferroic branch can provide the peculiar seesaw type magntoelectricity. One candiate family, i.e., Ti-based TMTC's monolayer, is predicted to own tunable alterferroicity. More importantly, the alterferroic concept and design rules can be generally applied to broader covalent ferroics near the magnetic/polar bicritical boundary. Our work not only completes the family of ferroicity, but also provides an unconventional strategy to realize intrinsic and strong magnetoelectric effects for next-generation information devices.

\textit{Methods}.
DFT calculations are performed using the Vienna \textit{ab initio} Simulation Pack (VASP) \cite{Kresse1996}. The projector augmented wave (PAW) pseudopotentials are Ti\_sv, C, Si, Ge, Te, and Pb. The plane-wave cutoff energy is fixed as 450 eV for TiCTe$_3$, and 350 eV for all others. The exchange is treated using Perdew-Burke-Ernzerhof (PBE) parametrization of the generalized gradient approximation (GGA). An effective Hubbard $U_{\rm eff}=2.5$ eV is adopted as default for Ti's $3d$ orbitals using the Dudarev approach \cite{Dudarev1998}. The choice of $U$ value for Ti$^{3+}$ is carefully tested (see SI Appendix, Text1 and Fig.~S10 for more details). 
	
To simulate a monolayer, a $15$ \AA{} vacuum layer is added along the $c$-axis direction to avoid the interaction between two neighboring slices. For Brillouin zone sampling, a $\Gamma$-centered $7\times7\times1$ and $7\times4\times1$ Monkhorst-Pack $k$-mesh are adopted for the rhombic cell and rectangular cell, respectively. Both the in-plane lattice constants and atomic positions are fully optimized iteratively until the Hellmann-Feynman force on each atom and the energy were converged to $0.01$ eV/\AA{} and $10^{-6}$ eV, respectively. The AMPLMODES software is used to seek for the distorted modes \cite{Orobengoa2009,Perez-Mato2010}, and the PHONOPY code is used to calculate the phonon spectra \cite{Togo2015}. The polarization is calculated by the Berry phase method~\cite{King-Smith1993}. The virtual crystal approximation (VCA) is employed to investigate the effects of ionic substitution \cite{Eckhardt2014}. The crystal structures are drawn using VESTA \cite{Momma2011}.

The random-field pseudo-dipole model is developed from the random-field Ising model for phase separation \cite{Mayr:Prl}. The Hamiltonian can be expressed as:
\begin{equation}
		H=\sum_{<ij>}J_{ij}(p_i\otimes p_j)+\sum_i R_i|p_i|-E\sum_i p_i,
\end{equation}
where tri-state $p_i$ stands for the local state at site $i$ ($0$: Z-AFM; $\pm1$: ferroelectric state with polarization up/down); The first item denotes the coupling between nearest-neighbor sites ($\otimes$ is not the simple multiplication). $R$ denotes the local random field from quench disorder which prefers the Z-AFM ($R_i>0$) or ferroelectric state ($R_i<0$); $E$ is the external electric field along the $c$-axis. More details of this model and Monte Carlo simulation can be found in SI Appendix.

\begin{acknowledgments}
	We are grateful to Jun Chen, Ning Ding, Churen Gui, Hai-Peng You, and Dr. Jun-Jie Zhang for helpful discussions. Work was supported by National Natural Science Foundation of China (Grant No. 11834002). We also thank the Big Data Computing Center of Southeast University for support of computational resources.
\end{acknowledgments}

\bibliography{pnas-reference}

\end{document}